\documentclass{article}
\usepackage{spconf,amsmath,graphicx}
\usepackage[table]{xcolor}
\usepackage{cite}
\usepackage{booktabs}
\usepackage{hyperref}
\usepackage{ dsfont }
\usepackage{float}

\definecolor{light-gray}{gray}{0.8}
\title{Ensemble knowledge distillation of self-supervised speech models}
\name{
\begin{tabular}{@{}c@{}}
Kuan-Po Huang$^{\star\dagger}$ \qquad 
Tzu-hsun Feng$^{\star}$ \qquad 
\thanks{$^\star$Equal contribution.}
Yu-Kuan Fu \qquad 
Tsu-Yuan Hsu\qquad\\
Po-Chieh Yen \qquad 
Wei-Cheng Tseng \qquad 
Kai-Wei Chang \qquad 
Hung-yi Lee
\end{tabular}}
\address{
$^{all}$College of Electrical Engineering and Computer Science, National Taiwan University\\
$^{\dagger}$ASUS Intelligent Cloud Services\\
\normalsize{\{f09922005, r10942095, r11942083, b08201047, b08901198, r09942094, f09921048, hungyilee\}@ntu.edu.tw}}

\begin{document}
\ninept
\maketitle
\begin{abstract}
Distilled self-supervised models have shown competitive performance and efficiency in recent years. However, there is a lack of experience in jointly distilling multiple self-supervised speech models. In our work, we performed Ensemble Knowledge Distillation (EKD) on various self-supervised speech models such as HuBERT, RobustHuBERT, and WavLM. We tried two different aggregation techniques, layerwise-average and layerwise-concatenation, to the representations of different teacher models and found that the former was more effective. On top of that, we proposed a multiple prediction head method for student models to predict different layer outputs of multiple teacher models simultaneously. The experimental results show that our method improves the performance of the distilled models on four downstream speech processing tasks, Phoneme Recognition, Speaker Identification, Emotion Recognition, and Automatic Speech Recognition in the hidden-set track of the SUPERB benchmark.
\end{abstract}
\begin{keywords}
Self-supervised Learning, Ensemble Knowledge Distillation, SUPERB, Distortions
\end{keywords}
\section{Introduction}
\label{sec:intro}

Recently, adopting self-supervised learned (SSL) models~\cite{mohamed2022self} has become a trend in speech processing. By leveraging a large amount of unlabeled data, \cite{baevski2020wav2vec, hsu2021hubert, chen2022wavlm} were able to achieve great performance on a variety of downstream speech processing tasks. As well as that, there is also some work focusing on the domain-shift problem of SSL models. A common out-of-domain scenario occurs when the testing data contains noises unseen during training. To overcome this problem, \cite{wang2022improving, wang2022wav2vec, huang22b_interspeech, chen2022wavlm} proposed various kinds of methods to enable models to produce robust representations of distorted speech. 

Despite the fact that SSL models are able to provide useful representations for various downstream speech processing tasks, these large models are wide and deep, making them inefficient for on-device speech applications. A straightforward method to reduce model size is to distill one of these models to obtain a compressed model with comparable performance. DistilHuBERT~\cite{chang2022distilhubert}, a small student model designed by reducing the depth of HuBERT~\cite{hsu2021hubert}, is trained by knowledge distillation with limited performance degradation compared to the HuBERT teacher model. Based on DistilHuBERT, some previous works~\cite{lee22p_interspeech, ashihara22_interspeech} studied the variant of the student models regarding the layer width and model depth, while some others~\cite{RobustDistilHuBERT} developed methods for enhancing the noise-robustness of knowledge-distilled models.




However, these models only leverage the advantages of a single SSL model. An intuitive way to combine knowledge of different models is to utilize the representations of different distilled models during downstream training. This method is inapplicable since it requires additional parameters and also degrades performance in some cases.
To leverage the strengths of each model while constraining the model size, a reasonable method is to adopt multiple SSL speech models as teachers during the knowledge distillation process. We hypothesize that having student models learning from multiple teachers helps acquire knowledge in a more general aspect. For example, distilling a model with high performance for clean speech and a model being robust to noise may result in a student model that performs well in both clean and noisy environments.

Though incorporating the concept of model ensembling into the teacher-student learning framework sounds intuitive, the aggregation method of the representations of each layer from different teacher models should be carefully considered. 
Recently, previous work showed that training speech recognition models with the concatenation of HuBERT~\cite{hsu2021hubert} and WavLM~\cite{chen2022wavlm} representations benefit speech recognition~\cite{arunkumar22b_interspeech}. Unfortunately, this is not the case in a knowledge distillation scheme. We found that concatenating is less effective than averaging the representations of different models. On top of that, there are existing works that ensembled multiple supervised trained neural networks~\cite{chebotar16_interspeech, gao2021distilling, wu2021one, wu2022unified} to improve downstream tasks. However, some of their methods rely on downstream results to decide the weights in the weighted-sum process of different model outputs in order to achieve the best performance. For SSL speech models, determining the weights for different model outputs according to various kinds of speech processing tasks cannot be easily achieved and is not reasonable since the prior knowledge of the types of downstream tasks should remain unknown during the pre-training stage. Different from previous work, we proposed to utilize multiple sets of prediction heads to predict different hidden layer outputs of multiple teacher models during knowledge distillation. 

Overall, we proposed to perform Ensemble Knowledge Distillation (EKD) to multiple SSL speech models to improve the performance of distilled models on different downstream speech processing tasks. Instead of averaging or concatenating the representations of different teacher models, we found that predicting each teacher model output with individual sets of prediction heads yields the best performance on four downstream tasks. Surprisingly, models trained with our proposed methods even improved downstream performance in noisy environments unseen during training. With our proposed method, it is no longer required to calculate a weighted sum of the representations of different teacher models and gives an insight into how ensemble learning can be conducted on SSL models. Most importantly, our proposed EKD method is downstream-independent, and does not have to be re-trained whenever there are new downstream tasks involved.

\begin{figure*}[ht]
\centering  
    \includegraphics[width=12.8cm]{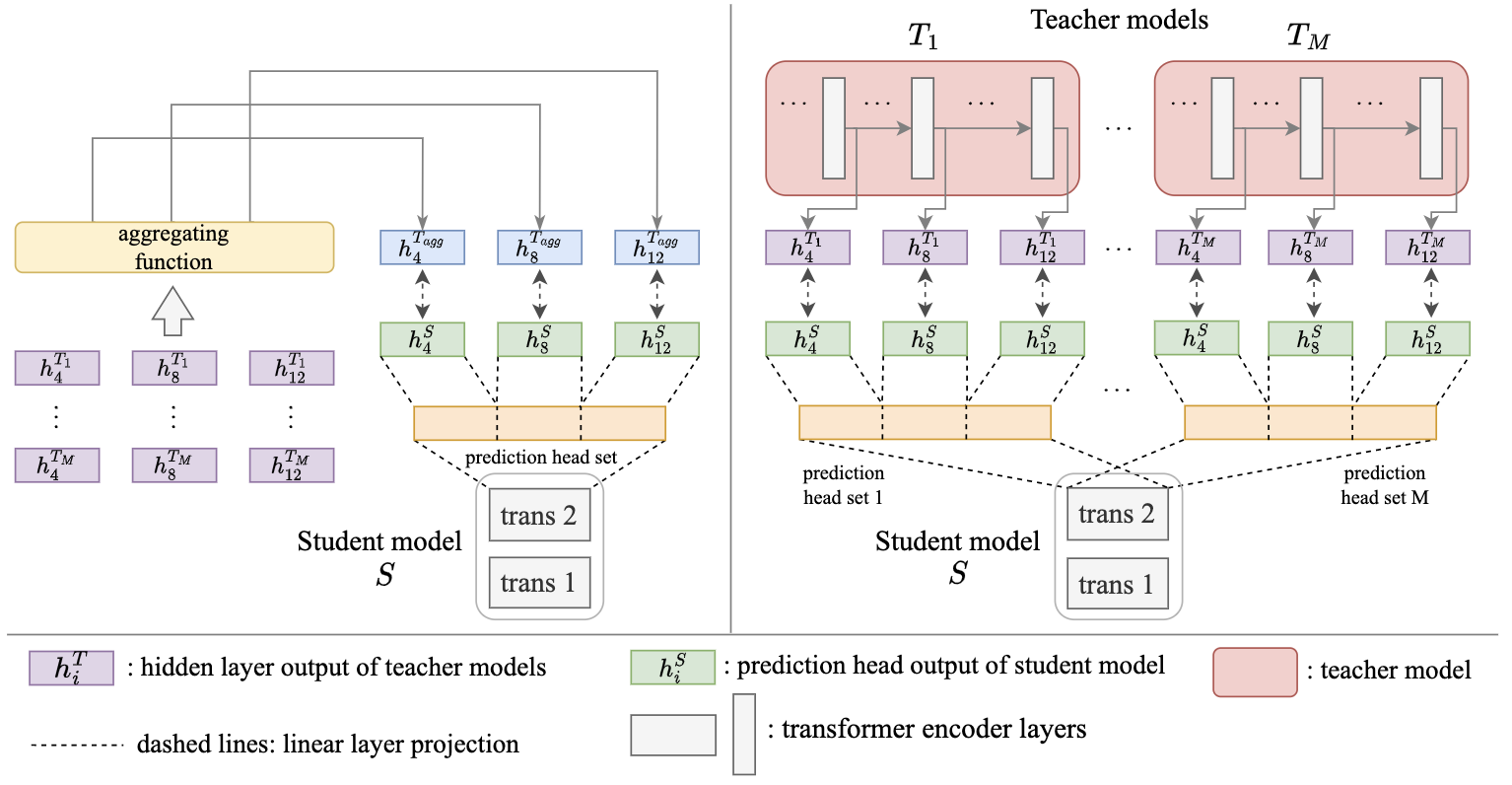}
    \vspace*{-4mm}
    \caption{Illustrations of EKD. The illustration on the left shows how EKD is performed with the aggregation of representations from multiple teacher models. The illustration on the right illustrates how EKD is performed with multiple sets of prediction heads.}
    \label{fig:pred_head}
\end{figure*}

\section{Ensemble Knowledge distillation (EKD)}
\subsection{Knowledge distillation}
Performing knowledge distillation refers to the process of distilling a teacher model $T$ into a smaller version. This is done by having a smaller student network $S$ to learn from the output of the teacher model. In DistilHuBERT~\cite{chang2022distilhubert}, the last layer output $z \in \mathds{R}^{t \times D_S}$ of the student model is transformed by prediction heads $Pr_i \in \mathds{R}^{D_S \times D_T}$ into $h^S_i \in \mathds{R}^{t \times D_T}$ to predict the $i^{th}$ hidden layer output $h^T_i \in \mathds{R}^{t \times D_T}$ of the teacher model. The role of the prediction heads is to transform the hidden layer output of the student to the same dimension as the hidden layer output of the teacher. The notation $t$ denotes the number of timesteps, while $D_S$ and $D_T$ denote the feature dimension of $z$ and $h^T_i$, respectively. The objective of the teacher-student framework of DistilHuBERT consists of an $L1$ loss and a cosine similarity loss as shown in Eq. (\ref{distil_loss}), where cossim\footnote{The cosine similarity operation is calculated by averaging the cosine similarity of the two feature vectors for each timestep.} is the cosine similarity operation, and $\sigma$ is the sigmoid activation.

\begin{equation}
\label{distil_loss}
\begin{aligned}
    \mathcal{L}_i(h^S_i, h^T_i) &= \frac{1}{D_T}\|h^S_i -  h^T_i \|_1 - \log \sigma(\mbox{cossim}(h^S_i, h^T_i))\\
    \mathcal{L} &= \frac{1}{3}\sum_{i \in \{4, 8, 12\}} \mathcal{L}_i(h^S_i, h^T_i)
\end{aligned}
\end{equation}

\subsection{Ensemble of teacher models}
Given a list of different teacher models, \{$T_1, \cdots, T_M$\}, where $M$ is the number of teacher models, to enable a student model to learn the outputs of different teacher models, some techniques will be required to define the output relationship between the teacher and student models. To adopt the same setting of DistilHuBERT having only one set of prediction heads to predict some of the hidden layer outputs of the teacher model, we aggregate the representations extracted from different teachers in a layerwise-averaged or a layerwise-concatenated manner. The distillation framework with aggregation of teacher representations is demonstrated in the left illustration of Fig.~\ref{fig:pred_head}.
\subsubsection{Layerwise-averaged representations}
\textbf{Layerwise-averaged} representations are generated by calculating the mean of the $i^{th}$ hidden layer output $h_i^{T_m}$ of each teacher model to form layerwise-averaged teacher representations $\overline{h}_i^{T} \in \mathds{R}^{t\times D_T}$ shown in Eq. (\ref{mean}).
    \begin{equation}
    \label{mean}
        \overline{h}_i^{T} = \frac{1}{M} \sum_{m=1}^M h_i^{T_m}
    \end{equation}
    For knowledge distillation, the objective for the student model to learn layerwise-averaged representations is shown in Eq. (\ref{distil_avg}).
    \begin{equation}
        \label{distil_avg}
        \mathcal{L}_{avg} = 
        \frac{1}{3}\sum_{i \in \{4, 8, 12\}} \mathcal{L}_i(h^S_i, \overline{h}^T_i)
    \end{equation}
\subsubsection{Layerwise-concatenated representations}
\textbf{Layerwise-concatenated} representations are generated by concatenating the $i^{th}$ hidden layer output $h_i^{T}$ for each of the $m^{th}$ teacher model to form layerwise-concatenated teacher representations $\hat{h}_i^{T_m} \in \mathds{R}^{t\times D_T^\prime}$ as shown in Eq. (\ref{concat}), where $D_T^\prime = D_T*M$. 
    \begin{equation}
    \label{concat}
        \hat{h}_i^{T} = \{ h_i^{T_1} \| h_i^{T_2} \| \cdots \| h_i^{T_M} \}
    \end{equation}
    For knowledge distillation, the objective for the student model to learn layerwise-concatenated representations is shown in Eq. (\ref{distil_concat}). Note that the dimension of the prediction heads $Pr_i \in \mathds{R}^{D_S \times D_T^\prime}$ in this case are different from the previous cases since the dimension of the layerwise-concatenated representations is different.
    \begin{equation}
        \label{distil_concat}
        \mathcal{L}_{concat} = 
        \frac{1}{3}\sum_{i \in \{4, 8, 12\}} \mathcal{L}_i(h^S_i, \hat{h}^T_i)
    \end{equation}

\subsubsection{Multiple sets of prediction heads}
To keep the original form of the representations from each teacher model, we proposed to adopt multiple sets of prediction heads following the last transformer encoder layer of the student model. Under the single-teacher knowledge distillation scheme, only a set of prediction heads are required to predict different layers of a single teacher model. However, to simultaneously learn from multiple teacher models during knowledge distillation, multiple sets of prediction heads are needed. 

\textbf{Multiple sets of prediction heads} are adopted to aim at learning different teacher model outputs without having to aggregate the teacher representations. The illustration on the right in Fig.~\ref{fig:pred_head} demonstrates the architecture of the distillation framework containing multiple sets of prediction heads. The quantity of the sets is equal to the number of teacher models involved during EKD. The predictions for each teacher model are generated by transforming the last hidden layer output of the student model with separate sets of prediction heads.
The objective of EKD when adopting different sets of prediction heads for each teacher model is shown in Eq. (\ref{eq:mp_loss}). $h_i^{S_m}$ is the prediction of the $m^{th}$ set of prediction heads trying to predict the hidden layer output $h_i^{T_m}$ of the $m^{th}$ teacher model.

\begin{equation}
    \label{eq:mp_loss}
    \mathcal{L}_{multi-pred} = 
    \frac{1}{3}\cdot\frac{1}{M}\sum_{m=1}^M \sum_{i \in \{4, 8, 12\}} \mathcal{L}_i(h^{S_m}_i, h^{T_m}_i)
\end{equation}

\section{Experimental setup}
\subsection{SUPERB hidden-set track}
From downstream evaluation, we report the results of tasks Phoneme Recognition (\textbf{PR}), Speaker Identification (\textbf{SID}), Emotion Recognition (\textbf{ER}), and Automatic Speech Recognition (\textbf{ASR}) in the hidden-set track of SUPERB challenge~\cite{SUPERBchallenge}.
This challenge aims to benchmark the generalizability of SSL speech models by adapting them to diverse speech processing tasks with lightweight downstream models appended.
The datasets in the hidden-set track are all newly created by the challenge organizers and stand unseen for each task. 
For downstream training, this challenge allows a weighted sum of representations extracted from the hidden layers to serve as the input of the downstream models. 

Following the same fashion as the challenge, we calculate an overall score with respect to predefined reference values\footnote{Details for calculating the overall score are shown in \href{https://superbbenchmark.org/challenge-slt2022/metrics}{https://superbbenchmark.org/challenge-slt2022/metrics}.}.
Downstream performance rankings are ranked according to this score.



\subsection{Noisy data}
Teacher models in our work are pre-trained with various kinds of noisy speech data. During the downstream evaluation, we also tested the performance with speech utterances containing background noises. We elaborate on different noise corpora in the following.

\textbf{Musan}~\cite{snyder2015musan} is an audio corpus containing audio samples of music, speech, and noises. The noise samples originate from the Freesound and Sound Bible database.

\textbf{WHAM!}~\cite{wichern2019wham} is a dataset containing real ambient noise samples recorded in the San Francisco Bay Area. It is originally used to simulate noisy environments of overlapped speech.

\textbf{DNS}~\cite{dubey2022icassp} is a  challenge that provides noise samples originating from Audio Set~\cite{gemmeke2017audio}, Freesound, and DEMAND~\cite{thiemann2013diverse} database. Synthetic room impulse responses are also released for augmenting speech samples to simulate a reverberated environment. Noise samples provided in DNS are involved in the augmentation process in the pre-training stage of WavLM~\cite{chen2022wavlm}.

\textbf{CHiME3}~\cite{barker2015third} refers to the third CHiME Challenge targeting speech recognition under real-world scenarios. The purpose of adopting CHiME3 is to construct a domain mismatch scenario during testing. Noises in this dataset are not involved during the pre-training stage of any of the self-supervised teacher models included in this work. For some speech processing tasks, it is difficult to find a corresponding noisy corpora. A simple way to obtain noisy testing data for downstream tasks is to add background noises to speech. Since we do not have access to the testing data in the hidden set, we consulted the SUPERB hidden-set committee and got permission to access the results of the testing sets of some tasks containing real-world background noises provided by the CHiME3 dataset. 
For Automatic Speech Recognition (\textbf{ASR}), since CHiME3 already provides a testing set for speech recognition recorded in real-world environments, we also report the results of this testing set (denoted as chime-real) in Table~\ref{tab:results}.

\subsection{Teacher models}
In our work, there are three different SSL teacher models involved.

\textbf{HuBERT}~\cite{hsu2021hubert} \textbf{(HB)}, an abbreviation of Hidden-Unit BERT, is a self-supervised speech model trained to predict clustered features with masked inputs. In our experiments, we adopt the base variant pre-trained with 960 hours of LibriSpeech~\cite{panayotov2015librispeech}.

\textbf{RobustHuBERT}~\cite{RobustDistilHuBERT} \textbf{(RHB)} is the distortion-robust version of HuBERT obtained by performing domain-adaptive pre-training~\cite{RobustDistilHuBERT} (DAPT) to HuBERT. The DAPT data originates from LibriSpeech but is distorted with background noises sampled from Musan, WHAM!, or Gaussian noise and augmented by applying reverberation, band rejection, or pitch-shifting.

\textbf{WavLM}~\cite{chen2022wavlm} \textbf{(WL)}, a similar model compared to HuBERT, achieved state-of-the-art performance by including gated relative position bias in the transformer-based architecture and augmenting input data into noisy or overlapped speech. In our experiment, we adopt the base+ variant pre-trained with 94k hours of speech.


\subsection{Training details}
For knowledge distillation\footnote{Code modified from \href{https://github.com/s3prl/s3prl}{https://github.com/s3prl/s3prl}.}, 960 hours of LibriSpeech are used for pre-training. Though some teacher models are robust to distortions, no additional distortions are added to the pre-training data for EKD.
The training hyper-parameters are similar to the original DistilHuBERT. The student network architecture is the same as DistilHuBERT except for the prediction heads under the multiple prediction head setting. To evaluate the models trained with EKD, we trained distilled versions of HuBERT, RobustHuBERT, and WavLM for comparison. For each model trained with EKD, their baselines are the single-teacher distilled versions of each teacher model involved in EKD.
For downstream speech processing, the data configuration and hyper-parameters for training follow the SUPERB hidden-set track. Prediction heads are discarded during this stage to reduce parameter usage. The training and evaluation process were conducted by the SUPERB hidden-set committee.

\section{Results}
\begin{table*}[t]
\setlength\tabcolsep{2.5pt}
\caption{Evaluation results on clean testing sets and testing sets with CHiME3 background noise. DT. is the abbreviation of Distil. The column ``method'' specifies whether the hidden layer outputs of the teacher are layerwise-averaged (avg.), layerwise-concatenated (concat.), or predicted with multiple sets of prediction heads (multi. pred.). The baselines of our proposed models are the individually distilled versions of the teacher models involved in the ensembling process. The values that outperform their corresponding baselines are marked gray. The best performance values of each testing set are marked bold. The rank of each model represents the overall performance on the clean set of the four downstream speech processing tasks.}
\centering
\begin{tabular}{cl|cc|cc|cc|cc|ccc|c}
\toprule
& & & & \multicolumn{2}{c|}{PR (PER ↓ )} & \multicolumn{2}{c|}{SID (Acc ↑ )} & \multicolumn{2}{c|}{ER (Acc ↑ )} & \multicolumn{3}{c|}{ASR (WER ↓ )} & \\
& & method & \# para. & clean  & chime   & clean  & chime  & clean & chime  & clean  & chime & chime-real & rank\\
\midrule
\midrule
(a) & DT. HB (baseline)~\cite{chang2022distilhubert}   & -    & 23M    & 35.50  & 42.70  & 74.33  & 71.25 & 54.67 & 53.85 & 65.64  & 74.05  & 74.21 & 10  \\
(b) &DT. WL (baseline) & -    & 23M    & 32.00    & 38.83 & 76.5   & 72.75 & 54.95 & 54.67 & 61.76  & 68.41  & 69.16 & 7  \\
\midrule
(c) & DT. HB \& WL    & avg.    & 23M    & 33.14 & 40.67 & \cellcolor{light-gray}\textbf{\emph{77.83}} & \cellcolor{light-gray}\textbf{\emph{73.92}}  & \cellcolor{light-gray}\emph{55.22}  & 54.67 & 61.85  & 69.75  & 70.94 & 5  \\
(d) & DT. HB \& WL    & concat. & 23M    & 33.95 & 41.51 & 75.83  & 71.75 & 54.67 & \cellcolor{light-gray}\emph{54.95}  & 62.17  & 70.36  & 69.47 & 8  \\
(e) & DT. HB \& WL    & multi. pred.   & 23M    & 32.46 & 39.57 & 75.00 & 72.50  & \cellcolor{light-gray}\emph{57.69}  & \cellcolor{light-gray}\textbf{\emph{56.59}}  & \cellcolor{light-gray}\emph{60.81} & 69.10   & \cellcolor{light-gray}\emph{68.64} & 2 \\
\midrule
(f) & DT. RHB (baseline)~\cite{RobustDistilHuBERT}  & -  & 23M    & 32.95 & 38.18 & 74.67  & 71.25 & 53.85 & 51.10  & 62.55  & 68.32  & 69.15 & 9 \\
\midrule
(g) & DT. HB \& RHB &  multi. pred.   & 23M    & \cellcolor{light-gray}\emph{32.30}  & 39.50  & \cellcolor{light-gray}\emph{76.17} & \cellcolor{light-gray}\textbf{\emph{73.92}}  & \cellcolor{light-gray}\emph{56.87}  & \cellcolor{light-gray}\emph{56.04}  & \cellcolor{light-gray}\emph{60.94} & \cellcolor{light-gray}\emph{67.80}  & \cellcolor{light-gray}\emph{69.12} & 4  \\
(h) & DT. RHB \& WL  &  multi. pred.   & 23M    & \cellcolor{light-gray}\textbf{\emph{31.02}}  & \cellcolor{light-gray}\textbf{\emph{37.33}}  & 75.17  &   70.58    & \cellcolor{light-gray}\emph{56.59}  & 53.30 & \cellcolor{light-gray}\emph{59.48} & \cellcolor{light-gray}\textbf{\emph{66.27}} & \cellcolor{light-gray}\textbf{\emph{65.21}} & 3\\
(i) & DT. HB \& RHB \& WL &  multi. pred.   & 23M    & \cellcolor{light-gray}\emph{31.14}  & \cellcolor{light-gray}\emph{38.06}  & 75.75  & 71.50 & \cellcolor{light-gray}\textbf{\emph{59.07}}  & \cellcolor{light-gray}\emph{56.32}  & \cellcolor{light-gray}\textbf{\emph{59.37}} & \cellcolor{light-gray}\emph{66.50}  & \cellcolor{light-gray}\emph{68.86} & 1 \\
\midrule
(j) & DT. HB \& DT. RHB \& DT. WL & - & 70M & 32.96	& 38.96 &	74.25 &	70.92 & 56.59 & 52.75 & 62.40 & 69.03 & 68.62& 6\\
\bottomrule
\end{tabular}

\label{tab:results}
\end{table*}
\subsection{Different aggregation methods}
In Table~\ref{tab:results}, (c)(d)(e) are models trained with EKD by having HuBERT and WavLM as their teachers. By comparing models (c) and (d), we discover that aggregating representations of the teacher in a layerwise-average manner tends to yield better performance than concatenating the representations. This may be due to the excessive length of representations being difficult to learn. 

By comparing model (c) to baseline models (a) and (b), we see that averaging the representations of HuBERT and WavLM improves the performance of \textbf{SID} and \textbf{ER} on the clean set, but degrades the performance of \textbf{PR} and \textbf{ASR}. This may indicate that the averaged representations from different models are not useful for content-based tasks.

\subsection{Multiple prediction heads}
In Table~\ref{tab:results}, we observe that model (e) outperforms models (c) and (d) on every task except for \textbf{SID}. This implies that having each set of prediction heads predict different teacher models is a better method compared to aggregating representations of the teacher in most cases. 


\subsection{Ensemble of distilled models}
The downstream results of model (j) are trained with the representations extracted from models (a)(b) and (f). From the rank, it is clear that utilizing representations of different distilled models improves downstream performance. However, the more distilled models used, the more parameters needed.

By comparing models (j) and (i), we discover that student models trained with EKD yield better performance than merely adopting representations of multiple individually-distilled student models. This is an inspiring result since the model trained with our proposed EKD method not only outperforms an ensemble of three distilled models but also requires only one-third of parameters.

\subsection{Different combinations of teacher models}
Models (e)(g)(h)(i) in Table~\ref{tab:results} are all trained by performing EKD with multiple sets of prediction heads and have the same training configurations but with different combinations of teacher models.

Model (g), trained to distill knowledge from HuBERT and RobustHuBERT, outperforms baseline models (a) and (f) on every testing set except for the noisy testing set of \textbf{PR}. This suggests that models distilled from a teacher model and its domain-adaptive pre-trained version not only gain robustness to noises but also gain improvement in clean environments for most of the tasks. Performing EKD shows great potential in this case for not degrading performance under clean and noisy settings. 

Among models (e)(g)(h)(i), model (i) performs the best in overall performance on the clean testing sets. Since ensembling more models does not increase model parameters as long as the prediction heads are not used during downstream training, it is worth trying more combinations of high-performance teacher models for EKD.

\section{Conclusion}
In our work, we conclude that performing Ensemble Knowledge Distillation to SSL speech models has the potential for improving model performance with restricted size in both clean and noisy environments. Having each teacher model predicted with separate sets of prediction heads is the best method for student models during knowledge distillation. Our method is also able to enhance the robustness of compressed models by jointly distilling teacher models while some teachers are not robust. In the future, we will try to ensemble more models to obtain better generalizability for small and efficient SSL speech models, and also integrate other noise-robust techniques into the EKD process.
\bibliographystyle{IEEEbib}
\bibliography{strings,refs}

\end{document}